\documentclass[aps,prl,reprint,notitlepage,showpacs,nofootinbib,superscriptaddress,twocolumn%,raggedbottom
,nofloats,preprintnumbers]{revtex4-1}
\usepackage{graphicx}
\usepackage[utf8]{inputenc} 
\usepackage{amsmath}
\usepackage{amssymb}
\usepackage{ulem}
\usepackage{float}
\usepackage{comment}
\usepackage{slashed}

\usepackage[T1]{fontenc} % if needed
\usepackage{parskip}
\usepackage{bm}         % for bold math characters in a math environment
\usepackage{array}
\usepackage{multirow}   % analogue to \multicolumn
\usepackage{booktabs}   % for fancy looking tables

\usepackage{textcomp}
\usepackage{amssymb}
\usepackage{ulem}
\usepackage{cancel}

\usepackage{enumerate}
\usepackage{caption}
\usepackage{subcaption}
\usepackage{ragged2e}
 \DeclareCaptionJustification{justified}{\justifying}
\usepackage{lineno}
\usepackage{amsmath}
\usepackage{amssymb}
\usepackage{indentfirst}
\usepackage[usenames,dvipsnames]{xcolor}
\usepackage[colorlinks=true,citecolor=darkred,urlcolor=darkred, pdfborder={0 0 0}]{hyperref}
\definecolor{darkred}{rgb}{0.6,0,0}

\definecolor{linkcolor}{rgb}{0,0,0.5}
\newcommand {\ignore}[1]{}

\definecolor{bostonuniversityred}{rgb}{0.8, 0.0, 0.0}

\def\gsim{\raise0.3ex\hbox{$\;>$\kern-0.75em\raise-1.1ex\hbox{$\sim\;$}}}
\def\lsim{\raise0.3ex\hbox{$\;<$\kern-0.75em\raise-1.1ex\hbox{$\sim\;$}}}

\usepackage{wrapfig}

\usepackage{soul}
\definecolor{mightnightblue}{RGB}{25,25,112}

\definecolor{brown}{rgb}{0.59, 0.29, 0.0}

\def\21{$\mathrm{SU(2)_L \otimes U(1)_Y}$}

\begin{document}

\title{Absorption of Fermionic Dark Matter in the PICO-60 C$_{3}$F$_{8}$ Bubble Chamber}

\author{E.~Adams}
\affiliation{Department of Physics, Queen's University, Kingston, K7L 3N6, Canada}

\author{B.~Ali}
\affiliation{Institute of Experimental and Applied Physics, Czech Technical University in Prague, Prague, Cz-12800, Czech Republic}

\author{R. Anderson-Dornan}
\email[Corresponding: ]{rajan.anderson@gmail.com}
\affiliation{Instituto de F\'{\i}sica, Universidad Nacional Aut\'onoma de M\'exico, A.P. 20-364, Ciudad de M\'exico 01000, M\'exico.}

\author{I.~J.~Arnquist}
\affiliation{Pacific Northwest National Laboratory, Richland, Washington 99354, USA}

\author{M.~Bai}
\affiliation{Department of Physics, Queen's University, Kingston, K7L 3N6, Canada}

\author{D.~Baxter}
\affiliation{Fermi National Accelerator Laboratory, Batavia, Illinois 60510, USA}

\author{E.~Behnke}
\affiliation{Department of Physics, Indiana University South Bend, South Bend, Indiana 46634, USA}

\author{B.~Broerman}
\affiliation{Department of Physics, Queen's University, Kingston, K7L 3N6, Canada}

\author{C.~J.~Chen}
\affiliation{
Department of Physics and Astronomy, Northwestern University, Evanston, Illinois 60208, USA}

\author{K.~Clark}
\affiliation{Department of Physics, Queen's University, Kingston, K7L 3N6, Canada}

\author{J.~I.~Collar}
\affiliation{Enrico Fermi Institute, KICP, and Department of Physics,
University of Chicago, Chicago, Illinois 60637, USA}

\author{P.~S.~Cooper}
\affiliation{Fermi National Accelerator Laboratory, Batavia, Illinois 60510, USA}

\author{D.~Cranshaw}
\affiliation{Department of Physics, Queen's University, Kingston, K7L 3N6, Canada}

\author{C.~Cripe}
\affiliation{Department of Physics, Indiana University South Bend, South Bend, Indiana 46634, USA}

\author{M.~Crisler}
\affiliation{Fermi National Accelerator Laboratory, Batavia, Illinois 60510, USA}

\author{C.~E.~Dahl}
\affiliation{
Department of Physics and Astronomy, Northwestern University, Evanston, Illinois 60208, USA}
\affiliation{Fermi National Accelerator Laboratory, Batavia, Illinois 60510, USA}

\author{M.~Das}
\affiliation{High Energy Nuclear \& Particle Physics Division, Saha Institute of Nuclear Physics, Kolkata, India}

\author{S.~Das}
\affiliation{High Energy Nuclear \& Particle Physics Division, Saha Institute of Nuclear Physics, Kolkata, India}

\author{S.~Fallows}
\affiliation{Department of Physics, University of Alberta, Edmonton, T6G 2E1, Canada}

\author{J.~Farine}
\affiliation{School of Natural Sciences, Laurentian University, Sudbury, ON P3E 2C6, Canada}
\affiliation{SNOLAB, Lively, Ontario, P3Y 1N2, Canada}
\affiliation{Department of Physics, Carleton University, Ottawa, Ontario, K1S 5B6, Canada}

\author{R.~Filgas}
\affiliation{Institute of Experimental and Applied Physics, Czech Technical University in Prague, Prague, Cz-12800, Czech Republic}

\author{A. Garc\'{\i}a-Viltres}
\affiliation{Instituto de F\'{\i}sica, Universidad Nacional Aut\'onoma de M\'exico, A.P. 20-364, Ciudad de M\'exico 01000, M\'exico.}

\author{G.~Giroux}
\affiliation{Department of Physics, Queen's University, Kingston, K7L 3N6, Canada}

\author{O.~Harris}
\affiliation{Northeastern Illinois University, Chicago, Illinois 60625, USA}

\author{H.~Hawley-Herrera}
\affiliation{Department of Physics, Queen's University, Kingston, K7L 3N6, Canada}

\author{T.~Hillier}
\affiliation{School of Natural Sciences, Laurentian University, Sudbury, ON P3E 2C6, Canada}

\author{E.~W.~Hoppe}
\affiliation{Pacific Northwest National Laboratory, Richland, Washington 99354, USA}

\author{C.~M.~Jackson}
\affiliation{Pacific Northwest National Laboratory, Richland, Washington 99354, USA}

\author{M.~Jin}
\affiliation{
Department of Physics and Astronomy, Northwestern University, Evanston, Illinois 60208, USA}

\author{C.~B.~Krauss}
\affiliation{Department of Physics, University of Alberta, Edmonton, T6G 2E1, Canada}

\author{M.~Laurin}
\affiliation{D\'epartement de Physique, Universit\'e de Montr\'eal, Montr\'eal, H2V 0B3, Canada}

\author{I.~Lawson}
\affiliation{School of Natural Sciences, Laurentian University, Sudbury, ON P3E 2C6, Canada}
\affiliation{SNOLAB, Lively, Ontario, P3Y 1N2, Canada}

\author{A.~Leblanc}
\affiliation{School of Natural Sciences, Laurentian University, Sudbury, ON P3E 2C6, Canada}

\author{H.~Leng}
\affiliation{Materials Research Institute, Penn State, University Park, Pennsylvania 16802, USA}

\author{I.~Levine}
\affiliation{Department of Physics, Indiana University South Bend, South Bend, Indiana 46634, USA}

\author{C.~Licciardi}
\affiliation{School of Natural Sciences, Laurentian University, Sudbury, ON P3E 2C6, Canada}
\affiliation{SNOLAB, Lively, Ontario, P3Y 1N2, Canada}
\affiliation{Department of Physics, Carleton University, Ottawa, Ontario, K1S 5B6, Canada}

\author{W.~H.~Lippincott}
\affiliation{Fermi National Accelerator Laboratory, Batavia, Illinois 60510, USA}
\affiliation{Department of Physics, University of California Santa Barbara, Santa Barbara, California 93106, USA}

\author{Q.~Malin}
\affiliation{Department of Physics, University of Alberta, Edmonton, T6G 2E1, Canada}

\author{P.~Mitra}
\affiliation{Department of Physics, University of Alberta, Edmonton, T6G 2E1, Canada}

\author{V.~Monette}
\affiliation{D\'epartement de Physique, Universit\'e de Montr\'eal, Montr\'eal, H2V 0B3, Canada}

\author{C.~Moore}
\affiliation{Department of Physics, Queen's University, Kingston, K7L 3N6, Canada}

\author{R.~Neilson}
\affiliation{Department of Physics, Drexel University, Philadelphia, Pennsylvania 19104, USA}

\author{A.~J.~Noble}
\affiliation{Department of Physics, Queen's University, Kingston, K7L 3N6, Canada}

\author{H.~Nozard}
\affiliation{D\'epartement de Physique, Universit\'e de Montr\'eal, Montr\'eal, H2V 0B3, Canada}

\author{S.~Pal}
\affiliation{Department of Physics, University of Alberta, Edmonton, T6G 2E1, Canada}

\author{M.-C.~Piro}
\affiliation{Department of Physics, University of Alberta, Edmonton, T6G 2E1, Canada}

\author{S.~Priya}
\affiliation{Materials Research Institute, Penn State, University Park, Pennsylvania 16802, USA}

\author{C.~Rethmeier}
\affiliation{Department of Physics, University of Alberta, Edmonton, T6G 2E1, Canada}

\author{M.~Robert}
\affiliation{Department of Physics, Queen's University, Kingston, K7L 3N6, Canada}

\author{A.~E.~Robinson}
\affiliation{D\'epartement de Physique, Universit\'e de Montr\'eal, Montr\'eal, H2V 0B3, Canada}

\author{J.~Savoie}
\affiliation{D\'epartement de Physique, Universit\'e de Montr\'eal, Montr\'eal, H2V 0B3, Canada}

\author{S.~J.~Sekula}
\affiliation{School of Natural Sciences, Laurentian University, Sudbury, ON P3E 2C6, Canada}
\affiliation{SNOLAB, Lively, Ontario, P3Y 1N2, Canada}

\author{A.~Sonnenschein}
\affiliation{Fermi National Accelerator Laboratory, Batavia, Illinois 60510, USA}

\author{N.~Starinski}
\affiliation{D\'epartement de Physique, Universit\'e de Montr\'eal, Montr\'eal, H2V 0B3, Canada}

\author{I.~\v{S}tekl}
\affiliation{Institute of Experimental and Applied Physics, Czech Technical University in Prague, Prague, Cz-12800, Czech Republic}

\author{M.~Tripathi}
\affiliation{Enrico Fermi Institute, KICP, and Department of Physics,
University of Chicago, Chicago, Illinois 60637, USA}

\author{E.~V\'azquez-J\'auregui}
\email[Corresponding: ]{ericvj@fisica.unam.mx}
\affiliation{Instituto de F\'{\i}sica, Universidad Nacional Aut\'onoma de M\'exico, A.P. 20-364, Ciudad de M\'exico 01000, M\'exico.}

\author{U.~Wichoski}
\affiliation{School of Natural Sciences, Laurentian University, Sudbury, ON P3E 2C6, Canada}
\affiliation{SNOLAB, Lively, Ontario, P3Y 1N2, Canada}
\affiliation{Department of Physics, Carleton University, Ottawa, Ontario, K1S 5B6, Canada}

\author{W.~Woodley}
\affiliation{Department of Physics, University of Alberta, Edmonton, T6G 2E1, Canada}
 
\author{V.~Zacek}
\affiliation{D\'epartement de Physique, Universit\'e de Montr\'eal, Montr\'eal, H2V 0B3, Canada}

\author{J.~Zhang}
\altaffiliation[now at ]{Argonne National Laboratory}
\affiliation{
Department of Physics and Astronomy, Northwestern University, Evanston, Illinois 60208, USA}

\collaboration{PICO Collaboration}
\noaffiliation

\date{\today}

\begin{abstract}

Fermionic dark matter absorption on nuclear targets via neutral current interactions is explored using a non-relativistic effective field theory framework. An analysis of data from the PICO-60 C$_{3}$F$_{8}$ bubble chamber sets leading constraints on spin-independent absorption for dark matter masses below 23 MeV/\textit{c}$^2$ and establishes the first limits on spin-dependent absorptive interactions. These results demonstrate the sensitivity of bubble chambers to low-mass dark matter and underscore the importance of absorption searches in expanding the parameter space of direct detection experiments.

\end{abstract}

\maketitle

%\tableofcontents

\section{\label{sec:introduction}Introduction}

There is compelling evidence that a significant fraction of the Universe's matter is non-baryonic~\cite{EINASTO1974,Ostrike1974,Ostriker:1973uit,bd0fcfe34889413099a774ad5bded97a,10.2307/45312,Bahcall:2013epa,SDSS:2007umu,10.1093/mnras/stt572,Bennett_2013,Ade:1530672}. Extensive searches for particle dark matter (DM) are being conducted using highly sensitive detectors in underground laboratories, which operate with energy thresholds on the order of keV in environments with extremely low background~\cite{PhysRevLett.127.261802,XENON:2018voc,HORN2015504,PhysRevLett.128.011801,Cooley_2010,Kluck_2020}. These experiments aim to detect nuclear recoils resulting from the interaction between dark matter particles and target nuclei. These detectors primarily target Weakly Interacting Massive Particles (WIMPs) and can probe cross sections as low as $10^{-48}$ cm$^2$ for spin-independent (SI; scalar) interactions~\cite{aalbers_dark_2024} and $10^{-41}$ cm$^2$ for spin-dependent (SD; axial-vector) interactions~\cite{PhysRevD.100.022001,aalbers_dark_2024}, particularly for dark matter masses in the range of 10–100 GeV/$\textit{c}^2$. In searches focusing on spin-independent couplings, constraints have been established under the assumption of coherent elastic scattering between dark matter and atomic nuclei.

As dark matter direct detection experiments approach the neutrino fog, the point where neutrino scattering becomes an irreducible background to the potential dark matter signal (though directional detection efforts like CYGNUS~\cite{vahsen2020cygnus} offer a promising path forward), without claiming discovery, interest is rising in exotic dark matter.  A vast unexplored parameter space exists for lower mass dark matter (sub-GeV), demanding further investigation.   Current experiments effectively constrain these masses via absorptive processes~\cite{Dror:2019onn,dror_absorption_2020,dror_absorption_2021,ge_revisiting_2022,PhysRevLett.129.161803,PhysRevLett.132.041001,PhysRevLett.129.221802,ELECTRONABSPANDAx}, successfully probing MeV-scale dark matter through spin-independent absorption. Recently, compelling motivation for dark matter at this scale has emerged from its potential to explain the anomalous ionization rate observed in the Central Molecular Zone (CMZ), a region of the Milky-Way Galaxy~\cite{PhysRevLett.134.101001}.

    To date, explorations have largely focused on this spin-independent channel, neglecting a wealth of other plausible interaction mechanisms. However, recent work has examined magnetic dipole-based absorption in the context of direct detection~\cite{li_dark_2022}, and there is a depth of literature around sterile neutrino dark matter, with recent studies exploring sterile neutrino transitions involving magnetic dipoles~\cite{beltran_probing_2024,zhang_probing_2023,brdar_neutrino_2023,miranda_low-energy_2021}. Expanding searches to encompass spin-dependent and other absorption channels is crucial for a comprehensive exploration of the light dark matter parameter space.

\section{\label{sec:inelasticDM} Absorption of Fermionic dark matter}
In this work, dark matter interactions with nuclear targets are investigated via the neutral current process
\begin{equation}
\chi + N \rightarrow \nu + N\,,
\end{equation}
which can be generically described by a set of four-fermion operators:
\begin{equation}
\mathcal{L}_{\text{int}} \propto \frac{1}{\Lambda^2} (\overline{\chi}\,\Gamma_i\,\nu)(\overline{N}\,\Gamma_j\,N)\,,
\end{equation}

\noindent  where $\Lambda$ is the effective energy scale, and $\Gamma_i=\left\{\mathbf{1}, \gamma_5, \gamma_\mu, \gamma_\mu \gamma_5, \sigma_{\mu \nu}\right\}$ represents the possible Lorentz invariant bilinears. Motivated by the original study by Dror \textit{et al.}~\cite{Dror:2019onn,dror_absorption_2020}, previous investigations have focused primarily on a spin-independent absorption mode, motivated by an ultraviolet (UV) model with a vector coupling within the four-fermion interaction framework. While many UV models can produce this rate, plausible alternative interaction models exist that yield substantially different rates and require careful investigation. To explore a more complete spectrum of dark matter interactions, a non-relativistic effective field theory (NREFT) approach has been adopted, which has long been used for more typical elastic dark matter-nuclear interactions~\cite{fitzpatrick_effective_2013}. Notably, a similar formalism has recently been developed for muon-to-electron transitions, which share kinematic similarities with the process considered in this work~\cite{haxton_effective_2024,haxton_nuclear-level_2022}.

For cold dark matter with a mass comparable to or lighter than the target, its momentum, $p_\chi$, is negligible. Meanwhile, for the neutrino—a relativistic particle—$E_{\nu} \approx p_{\nu}$. The momentum transfer then takes the following form:
\begin{equation}
q = \frac{m_N m_{\chi} + \frac{m_{\chi}^2}{2}}{m_N + m_{\chi}},
    \end{equation}

\noindent where $m_N$ is the target mass and $m_{\chi}$ is the DM mass. For $m_N \gg m_{\chi}$, $q \approx m_{\chi}$. Recoil energies on the scale of the dark matter mass indicate that the search is focused on low-mass dark matter candidates ($\sim$ MeV/c$^2$), and the outgoing ultra-relativistic particle will have momentum $p_{\nu} \approx m_{\chi}$. This fixes any relativistic degrees of freedom to non-relativistic ones, allowing a description of this process using the non-relativistic operators $\vec{S}_{\chi}$, $\vec{S}_{N}$, $i\vec{q}$, and $\vec{v}_{n_{avg}}$. The only difference between these operators and those from the elastic case is the velocity operator, which is replaced with the average nucleon velocity operator $\vec{v}_{n_{avg}} = \frac{v_{N_i} + v_{N_f}}{2}$ since it is not expected that the velocity of the incoming dark matter particle contributes meaningfully to the scattering. 

In the non-relativistic field theory, the dark matter component of the interaction can be treated separately from the nuclear part. By changing the velocity operator from one which denotes the velocity difference between the target and the dark matter to simply the average nucleon velocity, the dark matter part of the interaction is no longer velocity dependent and the nuclear responses provoked by each operator change. The recoil energy of the dark matter will also exhibit a unique peaked signature at $E_R \approx \frac{m_\chi^2}{2 m_N}$. This dark matter signal mimics that of a conventional WIMP, but during a threshold scan, an abrupt cutoff will be observed once the detector threshold surpasses the characteristic recoil energy. In the past, the PICO collaboration has used the WIMpy code~\cite{Kavanagh2015} to model photon-mediated interactions through the elastic NREFT~\cite{PhysRevD.106.042004}. The code has been modified to extend its applicability to absorptive interactions by incorporating the changes in velocity dependence and nuclear responses described above.

This letter centers on two particularly interesting operators within the NREFT. The first is $\mathcal{O}_1 = 1$, representing the leading-order spin-independent response. This operator corresponds to what other research groups have extensively investigated when studying spin-independent interactions. The second operator examined is $\mathcal{O}_4 = \vec{S}_\chi \cdot \vec{S}_N$,  representing the leading-order spin-dependent interaction and holding significant relevance for the magnetic dipole transitions where the $\mathcal{O}_{1}$ operator is not relevant. Although a multitude of effective operators can be considered, the matching process from UV or four-fermion weak effective theories to the NREFT often reveals that if certain operators are present with comparable coupling strengths, one operator tends to dominate. $\mathcal{O}_1$ is frequently this dominant operator. Following $\mathcal{O}_1$, the interaction mediated by $\mathcal{O}_4$ typically exhibits the next strongest response, underscoring the necessity of including spin-dependent interactions in this analysis. The NREFT framework employed here offers broad applicability to any cold dark matter-nucleus interaction where the dark matter's energy is at or below the nuclear energy scale. While $\mathcal{O}_4$ is crucial for understanding magnetic dipole transitions, it also emerges as the leading-order operator in a variety of other four-fermion interactions. The power of this framework lies in its capacity to systematically categorize and analyze a diverse set of potential dark matter interactions through a model-independent treatment that relies solely on the relevant kinematic variables.

\section{\label{sec:pico60}PICO-60 C$_{3}$F$_{8}$ bubble chamber and data analysis}

The PICO collaboration has operated multiple bubble chambers at the SNOLAB underground facility~\cite{Smith:2012fq} utilizing fluorocarbon fluids as the target medium. A key advantage of PICO’s bubble chambers is their strong sensitivity to nuclear recoils while simultaneously demonstrating a high level of rejection for electron recoil backgrounds.

These detectors consist of a high-purity synthetic fused silica vessel and stainless steel bellows, all enclosed within an stainless steel pressure vessel filled with hydraulic fluid. The inner chamber contains a fluorocarbon liquid (C$_{3}$F$_{8}$), while the pressure vessel is submerged in a water tank which provides shielding from external radiation and ensures thermal stability. Cameras continuously monitor the chambers, serving as both triggers and tools for position reconstruction by capturing images of bubble formation. Low-radioactivity piezoelectric transducers are attached to the silica vessel to record the acoustic signals generated during bubble formation, enabling the rejection of alpha decay backgrounds. To characterize detector performance, nuclear and electron recoil calibrations are performed {\it in-situ} using neutron sources (AmBe and Cf-252) and gamma sources (Co-60 and Ba-133)~\cite{PhysRevD.100.082006}, respectively. The detectors are operated such that a bubble forms only when energy exceeding a thermodynamic threshold is deposited within a critical radius—this is known as the Seitz threshold~\cite{1958_Seitz}. Bubble nucleation is therefore a threshold process governed by the Seitz model, which imposes a hard cutoff based on the thermodynamic conditions required for a phase transition.

The PICO-60 bubble chamber was filled with 52.2 kg of C$_{3}$F$_{8}$ and achieved exposures of 1167 kg-days at a 3.29-keV Seitz threshold and 1404 kg-days at a 2.45-keV Seitz threshold. These data were collected over two physics runs, conducted between November 2016 and January 2017~\cite{PhysRevLett.118.251301}, and between April and June 2017~\cite{PhysRevD.100.022001}, respectively. This experiment led to the most stringent direct detection constraints using fluorine on the WIMP-proton spin-dependent cross section, reaching a sensitivity of $2.5 \times 10^{-41}$ cm$^2$ for a 25 GeV/\textit{c}$^2$ WIMP~\cite{PhysRevD.100.022001}. Details on the experimental setup, data analysis, background estimates, and WIMP search results can be found in Refs.~\cite{PhysRevLett.118.251301,PhysRevD.100.022001}. In summary, three single-bubble events and five multi-bubble events were observed across the two runs, consistent with the expected neutron background ratio predicted by Monte Carlo simulations.

Exclusion limits were determined using efficiency curves derived from calibration data using various neutron sources, in several smaller bubble chambers~\cite{Durnford_2022}. Mono-energetic neutrons with energies of 50, 61, and 97 keV were generated via $^{51}$V(p,n)$^{51}$Cr reactions, while SbBe sources produced 24 keV neutrons, supplemented by AmBe neutron data~\cite{PhysRevD.100.022001}. This work follows the limit calculation method for SD and SI couplings previously published by the PICO collaboration~\cite{PhysRevD.100.022001}. Specifically, exclusion limits are determined via a Profile Likelihood Ratio (PLR) test~\cite{PLR_2011}.  Efficiency functions, initially derived from calibration data using the Markov Chain Monte Carlo (MCMC) as implemented in the emcee python package~\cite{Foreman_Mackey_2013,Durnford_2022,PhysRevD.106.122003}, provide the starting parameters for iterative MCMC optimization of DM sensitivities. This optimization maximizes a log-likelihood function using the aforementioned calibration data.  Iteration involves reseeding MCMC walkers based on DM sensitivity evaluations at ($2.45\pm 0.09$) and ($3.29\pm 0.09$) keV thresholds.  The result is a likelihood surface describing the DM sensitivity for each interaction type, threshold, DM mass, that contains information about the uncertainty of the nucleation efficiency.To incorporate the uncertainty in the Seitz threshold, each surface is convolved with a 2D Gaussian kernel, yielding likelihood surfaces that reflect both nucleation efficiency and thermodynamic uncertainties. The likelihood surface maxima then define the optimal couplings for each DM mass.

A test statistic is then constructed using the profile likelihood ratio test statistic, $q_\mu$. In order to obtain this test statistic, the following likelihood function is used,

\begin{equation}
    L(\mu) = L_{simple}(\mu) \cdot L_{NRE}(s), \label{eq:totLike}
\end{equation}

where $L_{simple}(\mu)$ represents a simple Poisson likelihood function, and $L_{NRE}(s)$ is obtained from the likelihood surfaces describing the uncertainties present in the DM-sensitivity due to the experimental uncertainties from the thermodynamic threshold and nucleation efficiency. Then, for a given DM mass and strength parameter ($\mu$), the likelihood function is maximized, assuming a maximally conservative background. The PLR is the ratio of this maximized likelihood to the global maximum likelihood across all cross sections (equivalent to the strength parameter in this case). If $\mu$ is above the best-fit cross section the test statistic is defined as $q_\mu = -2 \ln(\text{PLR})$, otherwise $q_\mu = 0$. Toy Monte Carlo datasets are generated for each DM mass and the cross section being examined. A point is excluded if its $q_\mu$ exceeds that of 90\% of the corresponding toy datasets.

The exclusion limits were calculated under the standard halo parametrization~\cite{LEWIN199687}. These calculations assume a local dark matter density of $\rho_D=0.3~{\mathrm{GeV\textit{c}^{-2}cm^{-3}}}$, along with the same astrophysical parameters used in prior analyses. The nuclear recoil energy window extends from the thermodynamic threshold up to 100 keV, with the upper bound conservatively chosen due to the lack of acoustic calibration for recoils above $\sim$100 keV. The reported limits remain consistent with the absence of a dark matter signal.

\section{\label{sec:results} Results}

The results can be presented using different parameterizations. One common approach is to set limits on the operator couplings. However, to facilitate a direct comparison with existing literature, and following the approach of Dror et al.~\cite{Dror:2019onn,dror_absorption_2020} and other collaborations, this work instead expresses constraints in terms of a generic cross section, assuming dark matter only couples through the operator of interest. It should be noted, however, that this parameterization may not be ideal for all interaction channels. This cross section is expressed as follows:
\begin{equation}
\sigma_{NC} = \frac{m_\chi^2}{4\pi\Lambda^4}.
\end{equation}

\begin{figure}[htpb!]
    \centering
    \includegraphics[width=\linewidth]{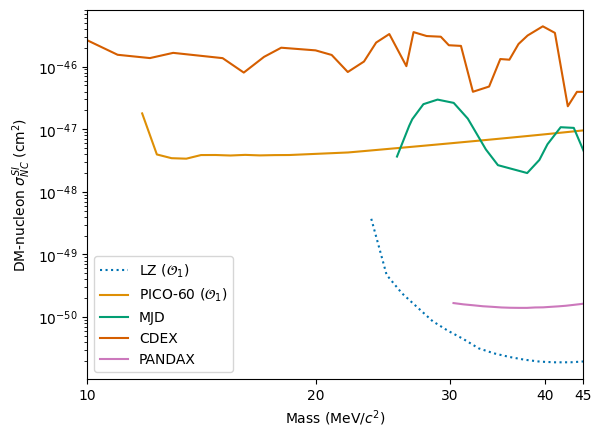}
    \caption{ Upper limits (90\% C. L.) of the fermionic DM absorption cross section as a function of the dark matter mass for the effective operator $\mathcal{O}_1$. The results from the PICO-60 C$_3$F$_8$ bubble chamber analysis are presented (orange). Additionally, results from the CDEX-10~\cite{PhysRevLett.129.221802} (red), PANDAX-4T~\cite{PhysRevLett.129.161803} (purple), and Majorana Demonstrator (MJD)~\cite{PhysRevLett.132.041001} (green) experiments are also shown. The sensitivity from the LZ experiment is also presented, by applying the NREFT framework to the latest dataset reported~\cite{aalbers_dark_2024} (blue dotted).}
    \label{fig:O1}
\end{figure}

Figure~\ref{fig:O1} presents the 90\% C. L. exclusion limits of the fermionic DM
absorption cross section for the effective operator $\mathcal{O}_1$ obtained from the analysis of the data from the PICO-60 C$_3$F$_8$ bubble chamber. This experiment sets leading limits on dark matter with masses below 23 MeV/c$^2$. Results from CDEX-10~\cite{PhysRevLett.129.221802}, PANDAX-4T~\cite{PhysRevLett.129.161803}, and Majorana Demonstrator~\cite{PhysRevLett.132.041001} are also shown. Additionally, the sensitivity obtained from the most recent LZ dataset~\cite{aalbers_dark_2024} is also presented, by applying the NREFT framework developed in this work. The sharp rise observed in the exclusion curves corresponds to the kinematic cutoffs imposed by the carbon and fluorine isotopes in the detector. Compared to idealized theoretical calculations, these features appear smoothed due to the treatment of uncertainties in the likelihood surfaces.

\begin{figure}[htpb!]
    \centering
    \includegraphics[width=\linewidth]{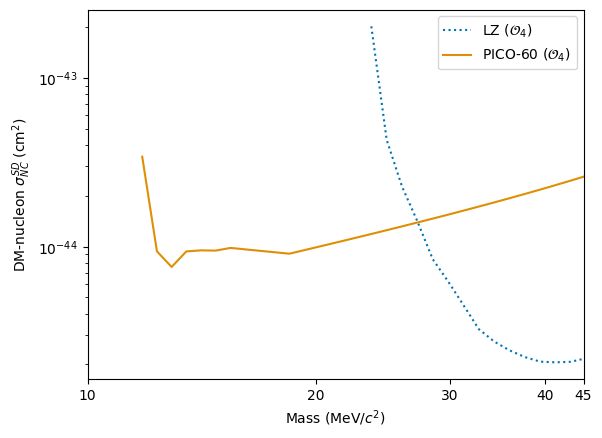}
    \caption{ Upper limits (90\% C. L.) of the fermionic DM absorption cross section as a function of the dark matter mass for the effective operator $\mathcal{O}_4$. Results from the PICO-60 C$_3$F$_8$ bubble chamber (orange) are shown. The sensitivity from the LZ experiment is also presented, by applying the NREFT framework to the latest dataset reported~\cite{aalbers_dark_2024} (blue dotted).}
    \label{fig:O4}
\end{figure}

The first limits (90\% C. L.) on fermionic DM absorption that couples to the nucleus through the leading order spin-dependent operator $\mathcal{O}_{4}$ are presented in Figure~\ref{fig:O4}. The expected sensitivity of the LZ experiment is also shown. Both figures present an important part of the parameter space, as they help bridge the sensitivity gap between keV-scale dark matter searches targeting electron interactions~\cite{dror_absorption_2021,ge_revisiting_2022} and higher MeV-scale dark matter searches using xenon targets.

Both cases highlight the advantage of using low-mass-number isotopes, as the recoil energy threshold scales inversely with the nuclear mass ($E_R \approx \frac{m_\chi^2}{2 m_N}$).

\section{\label{sec:conclusions} Conclusions and discussion}

Leveraging the kinematic similarities with muon-to-electron transitions, this Letter presents a comprehensive, model-independent framework for analyzing fermionic dark matter absorption onto nuclear targets via neutral current interactions. The non-relativistic effective field theory approach has been, for the first time, extended to the absorptive case, enabling systematic exploration of diverse interaction channels, moving beyond specific, often weakly-motivated, UV models.
This analysis of PICO-60 data demonstrates that bubble chambers offer exceptional sensitivity to low-mass (MeV-scale) fermionic dark matter. PICO-60 sets world-leading limits on spin-independent absorption for dark matter masses below 23 MeV/$c^2$. Furthermore, the first constraints on spin-dependent absorptive interactions are presented, which is crucial for scenarios like magnetic dipole-mediated absorption. Specifically, PICO-60 provides the most stringent limits on the spin-dependent operator $\mathcal{O}_4$ within the 12 to 27 MeV/$c^2$ range. The aforementioned mass ranges are of particular interest due to their potential to explain the anomalous ionization rate in the CMZ through dark matter annihilation, which may play a significant role in solving the origin of the 511 keV line emission. These results advance low-mass fermionic dark matter searches via NREFT operator constraints, opening a broad parameter space and emphasizing absorption as a key beyond WIMP discovery avenue. PICO bubble chambers continue probing dark matter scenarios with unique sensitivity. 

\section{Acknowledgements}
\begin{acknowledgments}

The PICO collaboration wishes to thank Eduardo Peinado for useful discussions and SNOLAB and its staff for support through underground space, logistical, and technical services. SNOLAB operations are supported by the Canada Foundation for Innovation (CFI) and the Province of Ontario Ministry of Research and Innovation, with underground access provided by Vale at the Creighton mine site. The authors wish to acknowledge the support of the Natural Sciences and Engineering Research Council of Canada (NSERC), CFI for funding, and the Arthur B. McDonald Canadian Astroparticle Physics Research Institute. The authors acknowledge that this work is supported by the National Science Foundation (NSF)
(Grants No. 0919526, No. 1506337, No. 1242637, and No. PHY-241165), by the U.S. Department of Energy (DOE) Office of Science, Office of High Energy Physics (Grants No. DE-SC0017815 and No. DE-SC-0012161), by the DOE Office of Science Graduate Student Research (SCGSR) award, by the Department of Atomic Energy (DAE), Government of India, under the Centre for AstroParticle Physics II project (CAPP-II) at the Saha Institute of Nuclear Physics (SINP), and Institutional support of Institute Experimental and Applied Physics (IEAP), Czech Technical University in Prague (CTU) (DKRVO). This work is also supported by DGAPA UNAM Grant No. PAPIIT IN105923, and Fundaci\'on Marcos Moshinsky. This work is partially supported by the Kavli Institute for Cosmological Physics at the University of Chicago through NSF Grants No. 1125897 and No. 1806722, and an endowment from the Kavli Foundation and its founder Fred Kavli. The authors also wish to acknowledge the support from Fermi National Accelerator Laboratory under Contract No. DEAC02-07CH11359, and from Pacific Northwest National Laboratory, which is operated by Battelle for the U.S. Department of Energy under Contract No. DE-AC05-76RL01830. The authors also thank Digital Research Alliance of Canada~\cite{ComputeCanada} for computational support. E. V.-J. is grateful for the support of PASPA-DGAPA, UNAM for a sabbatical leave.
%The work of D. Durnford is supported by the NSERC Canada Graduate Scholarships-Doctoral program (CGSD).
\end{acknowledgments}
\bibliographystyle{apsrev4-1}
\bibliography{apssamp}

\end{document}